\documentclass{article}
\usepackage{ijcai24}

\usepackage{times}
\usepackage{soul}
\usepackage{url}
\usepackage[hidelinks]{hyperref}
\usepackage[utf8]{inputenc}
\usepackage[small]{caption}
\usepackage{graphicx}
\usepackage{amsmath}
\usepackage{amsthm}
\usepackage{booktabs}
\usepackage{algorithm}
\usepackage{algorithmic}

\usepackage[switch]{lineno}
\usepackage{amssymb}

\usepackage{xr}
\makeatletter

\newcommand*{\addFileDependency}[1]{
\typeout{(#1)}
\@addtofilelist{#1}
\IfFileExists{#1}{}{\typeout{No file #1.}}
}\makeatother

\newcommand*{\myexternaldocument}[1]{%
\externaldocument{#1}%
\addFileDependency{#1.tex}%
\addFileDependency{#1.aux}%
}

\usepackage{lipsum}

\newcommand\blfootnote[1]{%
  \begingroup
  \renewcommand\thefootnote{}\footnote{#1}%
  \addtocounter{footnote}{-1}%
  \endgroup
}

\myexternaldocument{supplement}

\title{Enhancing Cooperation through Selective Interaction and Long-term Experiences in Multi-Agent Reinforcement Learning}

\author{
Tianyu Ren
\and
Xiao-Jun Zeng
\affiliations
University of Manchester
\emails
\{tianyu.ren, x.zeng\}@manchester.ac.uk
}

\begin{document}

\maketitle

\begin{abstract}
The significance of network structures in promoting group cooperation within social dilemmas has been widely recognized. Prior studies attribute this facilitation to the assortment of strategies driven by spatial interactions. Although reinforcement learning has been employed to investigate the impact of dynamic interaction on the evolution of cooperation, there remains a lack of understanding about how agents develop neighbour selection behaviours and the formation of strategic assortment within an explicit interaction structure. To address this, our study introduces a computational framework based on multi-agent reinforcement learning in the spatial Prisoner's Dilemma game. This framework allows agents to select dilemma strategies and interacting neighbours based on their long-term experiences, differing from existing research that relies on preset social norms or external incentives. By modelling each agent using two distinct Q-networks, we disentangle the coevolutionary dynamics between cooperation and interaction. The results indicate that long-term experience enables agents to develop the ability to identify non-cooperative neighbours and exhibit a preference for interaction with cooperative ones. This emergent self-organizing behaviour leads to the clustering of agents with similar strategies, thereby increasing network reciprocity and enhancing group cooperation. 
\end{abstract}

\section{Introduction}
\blfootnote{Proceedings of the Thirty-Third International Joint Conference on Artificial Intelligence (IJCAI-24)}
The emergence of cooperation is fundamental to human civilization and evident in various biological and multi-agent systems ~\cite{rand2013human,kramar2022negotiation}. However, maintaining cooperation in social dilemmas is challenging due to the conflict between individual interests and collective welfare~\cite{sigmund2010calculus}. A key focus in cooperation research involves understanding the conditions under which individuals favour altruism over self-interest, with reciprocity identified as a crucial element. In this context, reciprocity is examined through repeated interactions limited to direct neighbours due to physical or social constraints, where the behaviours of individuals are shaped by the actions of their counterparts~\cite{perc2013evolutionary}. Studies using evolutionary game theory (EGT) demonstrate that network structure can promote altruistic behaviour through spatial reciprocity,  clustering similar strategies and reducing exploitation by free-riders~\cite{santos2006cooperation,wang2013insight}. Furthermore, considering individuals' capability to interact with neighbours selectively, recent research emphasizes the significance of dynamic interaction mechanisms in the evolution of cooperation~\cite{su2022evolution,rand2011dynamic,sylwester2013reputation}. Such neighbour selection behaviours notably transform learning dynamics, thereby allowing agents to modify their interaction structures in response to evolving cooperative scenarios.

Despite advancements, explanations for the spontaneous emergence of cooperative behaviours and selective interactions remain limited. Traditional game theory and EGT emphasize social learning, where agents imitate nearby successful strategies~\cite{sigmund2010social,li2016changing}, overlooking learning through trial-and-error~\cite{metcalfe2017learning}. Additionally, empirical studies employing experience-based learning face challenges in developing enduring strategies, particularly due to the complexities encountered in large population iterations~\cite{mckee2023scaffolding}. To unravel the intricate coevolutionary dynamics of agent behaviours, there is an increasing interest in leveraging advanced deep reinforcement learning (RL) algorithms~\cite{perolat2017multi,du2023review,willis2023resolving}. These methods are not only used to examine agents' decision-making processes but also to investigate the emergence of their behaviours~\cite{koster2022spurious}.  However, RL agents typically optimize personal policies, which may lead to a reduction in global optimization~\cite {lowe2017multi}. Although studies introducing mechanisms like reputation~\cite{anastassacos2020partner} and moral rewards~\cite{tennant2023modeling} to address these issues, these specialized approaches exhibit restricted applicability. Therefore, it is crucial to have a deeper understanding of the learning dynamics of agent behaviours and encourage cooperation.

In this study, we construct a computational model using deep Q-learning (DQN)~\cite{mnih2015human} to explore how agents can simultaneously learn both interaction and dilemma strategies from their long-term experiences. These agents, modelled as artificial neural networks, learn behavioural policies and obtain rewards in a spatial Prisoner's Dilemma Game (PDG) setting within a multi-agent reinforcement learning (MARL) environment. Unlike previous studies that depended on predefined social norms or explicit external incentives, our approach highlights the significance of temporal factors and historical information in influencing agent decision-making. Initially, agents have no prior knowledge regarding the actions or game states, hindering their ability to assess the consequences of their actions and respond effectively.  Throughout the training process, they must learn the causality between their actions, observations, and rewards from local environment observation. To aid this learning process, we introduce a utility function that integrates self-learning and social learning, reflecting the interplay of personal preferences and the influence of others in shaping human behaviour.

The experimental results demonstrate that RL agents trained in our framework effectively differentiate between cooperative neighbours and those who are free-riders. Their preference for building connections with cooperators bolsters network reciprocity, contributing to the formation of strategy clusters in network-structured populations. This finding aligns with existing EGT research, emphasizing the importance of strategy assortment in promoting cooperation. Moreover, the trained agents achieve superior cooperation levels and greater average payoff compared to the EGT baseline. Further, we observe that increased efficiency in learning is correlated with the length of memory experiences. For a detailed comparison between our RL model and EGT approaches, refer to the Supplementary Information (SI) \ref{Supp:Hyperparameters}. 

Our work offers three key contributions.  Firstly,  it reveals the coevolutionary dynamics of cooperation and interaction strategies within a spatial PDG framework, demonstrating that RL agents can learn effective interaction mechanisms to enhance network reciprocity and cooperation. Secondly, it sheds light on how extensive long-term experiences positively influence group cooperation. Finally, the MARL training environment we developed sets the stage for future explorations into diverse aspects of pro-social cooperative behaviour.

\section{Related Works}
\subsection{Evolutionary Game Theory}
EGT is crucial for exploring the evolution of cooperation among self-interested individuals. It expands on traditional game theory by considering extended interactions and strategy dynamics, exploring the emergence and stability of cooperative behaviours. Notably, Nowak~\shortcite{nowak2006five} identifies five mechanisms central to understanding cooperation evolution. One vital aspect noted is the emergence of cooperation on network structures, wherein individuals predominantly interact with their immediate neighbours~\cite{perc2017statistical}. 
 
Inspired by the dynamic nature of social interactions, numerous studies have explored the coevolutionary dynamics of cooperation by integrating strategy evolution with network changes.  In this domain, the concept of network assortativity has been highlighted, revealing that agents with similar strategies often connect, thereby boosting group cooperation~\cite {tanimoto2013difference,ren2021evolutionary}. Research like Su et al.~\shortcite{su2022evolution} explores how cooperative strategies evolve and spread, particularly in unidirectional interactions, shedding light on shaping social interactions to promote cooperation. However, these often neglect the self-learning aspect of agents, which is crucial in real-life where directly copying strategies is impractical. To address this gap and accurately depict cooperation emergence, we applied the MARL framework, enabling agents to learn both dilemma and interaction strategies through environmental observation independently.
 
 \subsection{Human Experiments}
While EGT is essential for examining evolutionary trajectories and conditions favouring cooperation, incorporating empirical data brings psychological nuances to these models~\cite{kobis2019intuitive}, which focus on imposed interaction structures and psychological mechanisms beyond laboratory settings~\cite{rand2013human}. Surprisingly, behavioural studies indicate that participants in structured settings tend to randomly change strategies instead of copying higher-payoff neighbours~\cite{traulsen2010human}, disrupting the clustering process and rendering cooperation less advantageous. Addressing this, Rand et al.~\shortcite{rand2011dynamic} found that dynamic interactions enhance multilateral cooperation by encouraging links with cooperators over defectors, promoting strategy clustering. However, such laboratory experiments often encounter challenges in terms of scalability and struggle with complex, larger-scale, or long-term scenarios ~\cite{moffatt2009experimental}.

\subsection{Multi-agent Reinforcement Learning}
Recent RL applications also focus on understanding the emergence of cooperation by integrating spatial and temporal dynamics relevant to realistic scenarios~\cite{ren2023reputation,vinitsky2023learning,tennant2023modeling}, moving beyond traditional matrix games through the fusion of complex incentive structures~\cite{leibo2017multi,jaques2019social,abeywickrama2023emergence}. Studies have applied the intrinsic trial-and-error learning characteristic of RL to reformulate agent interaction strategies. For instance, Anastassacos et al.~\shortcite{anastassacos2020partner} demonstrate that RL agents can learn an interaction strategy akin to Tit-for-Tat when partner selection is present, aiding in the maintenance of cooperation. Meanwhile, McKee et al.~\shortcite{mckee2023scaffolding} used a graph neural network-based agent as a social planner, demonstrating the ability of deep RL to foster coordination and cooperation in a group.

Concurrently with our work, Ueshima~\shortcite{ueshima2023deconstructing} employed two distinct Q-networks for each agent, differentiating interaction and dilemma strategies. However, our approach varies as follows: (1) we extend the model to allow each agent to interact with four potential neighbours, unlike their paired interaction focus; (2) we incorporate environments with an explicit interaction structure, considering network reciprocity;  (3) they consider single-round observation input, while we take into account the agents' long-term experiences.

\section{Background}
\subsection{Prisoner's Dilemma Game}
The PDG is a fundamental paradigm in EGT~\cite{rapoport1965prisoner}, representing a typical decision-making scenario where agents balance individual benefits against collective welfare. Fundamentally, the PDG  is characterized as a symmetric matrix game, representing interactions between pairs of individuals within a population. Each participant faces a choice: to cooperate  ($C$), incurring a cost $c$ while providing a benefit $b$ to others, or to defect ($D$), avoiding the cost while exploiting those who cooperate (with $b > c > 0$). The corresponding payoff matrix can be summarized as
\begin{equation}
\mathcal{M}_p=
\left[\begin{array}{ccc}
R & S\\
T & P \\
\end{array}
\right]
\end{equation}
where mutual cooperation yields a reward $R=b-c$, while mutual defection result in $P=0$. Unilateral cooperation against a defector incurs a cost $S=c$, wherease the defector gains $T=b$. This payoff matrix forms four classical game structures~\cite{wang2015universal}: PD, chicken, harmony and stag hunt, each defined by specific payoff inequalities. In the PD, the conditions $T>R>P>S$ and $2R>T+S$ hold. Our model employs a weak PDG with $T=b$ $(1\le b \le 2)$, $R=1$, and $P=S=0$~\cite{nowak1993spatial}. Here, the parameter $b$ directly assesses the strength of the dilemma, given $c=0$. In a single-shot PDG, defection is the dominant strategy, leading to a Nash Equilibrium of defection, despite cooperation could yield a Pareto improvement. Considering that players often engage in repeated iterations with the same counterparts, the conventional PD can be extended to the iterated prisoner's dilemma (IPD) format. In this work, the dilemma strategy of agent $i$ at timestep $t$ is represented by a two-dimensional unit vector $a_{d_i}(t)$, with $a_{d_i}=[1,0]^T$ indicating cooperation and $a_{d_i}=[0,1]^T$ signifying defection.

\subsection{MARL Markov Game}
Within MARL, the IPD is conceptualized as a multi-agent extension of Markov decision processes (MDPs), termed partially observable general-sum Markov games~\cite{littman1994markov}. Here, agents have observations limited to their local environment. Formally, a $N$-player MDP is defined by the tuple $\mathcal{M}=\langle \mathcal{S},\mathcal{\{A}_i\}_{i\in{\mathcal{N}}},\mathcal{T},\gamma,\mathcal{R}\rangle$, where $\mathcal{S}$ is a set of joint states for all agents, $\mathcal{A}_{1}, \dots, \mathcal{A}_N$ represent joint actions, and $\mathcal{R}$ is the reward function. The function $\mathcal{O}: \mathcal{S}\times\{1\dots, N\}\to \mathbb{R}^d $ maps each player’s $d$-dimensional view of the state space. In a given state, each agent $i$ selects an action from $\mathcal{A}_i$, and the dynamics of MDP are determined by the stochastic transition function $\mathcal{T}: \mathcal{S} \times \mathcal{A}_1 \times \dots \times \mathcal{A}_N \to \triangle(\mathcal{S})$, where $\triangle(\mathcal{S})$ represents the set of discrete probability distributions over $\mathcal{S}$. The objective of each agent $i$ is to learn a policy $\pi_i:\mathcal{O}_i\to\triangle(A_i)$ to maximize its extrinsic reward $r_i(s,a^1,\dots,a~^N)$ based on the agent's own observation, simplified as $\pi(a^i|o^i)$. This optimization is conducted while adhering to the joint policy of all agents. The long-term $\gamma$-discounted payoff for agent $i$ under the joint policy $\overrightarrow{\pi} = (\pi_1, \dots, \pi_N)$ from an initial state $s_0$ can be defined as:
\begin{equation}
V_i^{\overrightarrow{\pi}}(s_0)=\mathbb{E}_{\overrightarrow{a_t}\sim \overrightarrow{\pi}(\mathcal{O}(s_t)),s_{t+1}\sim\mathcal{T}(s_a,\overrightarrow{a}_t)}[\sum^{T}_{t=0}\gamma^tr_i(s_t,\overrightarrow{a}_t)]
\end{equation}
where $\gamma\in[0,1]$ represents the temporal discount factor, and $T$ denotes the time horizon. Policies are optimized using trial-and-error interactions within the MARL environment to maximise cumulative long-term rewards.

\subsection{Deep Q-Network}
As an extension of Q-learning, DQN stands out as one of the most popular off-policy Deep RL algorithms, which utilizes an independent deep neural network to estimate Q-values~\cite{mnih2015human}. Departing from the traditional tabular representation for Q-values of state-action pairs, DQN utilizes a parametrized Q-function $Q_\theta(s, a)$ to approximate Q-values. Each Q-network is parameterized by $\theta$, representing the weights of the neural network. This approach incorporates the utilization of a replay memory buffer $\mathcal{D}$ to store past experiences and a target Q function $\overline{Q}$ to mitigate the risk of overestimating Q-values.  The learning process for the optimal action-value function $Q^*$ involves minimizing the loss.
\begin{equation}
\mathcal{L}_{\theta}=\mathbb{E}_{(s,a,r,s')\sim \mathcal{D}}[(r+\gamma\max_{a'
}\overline{Q}(s',a')-Q(s,a))^2].
\end{equation}

DQN can be directly extended to a multi-agent setting by having each agent $i$ learn an independent Q function denoted as $Q_i: \mathcal{O}_i \times \mathcal{A}_i \rightarrow \mathbb{R}$. In line with the traditional Q-learning approach, DQN also adopts an $\epsilon$-greedy policy to promote exploration. The policy for the $i$-th agent is parameterized as:

\begin{equation}  
\pi_i(s)=
\begin{cases}
\arg\max_{a_i\in \mathcal{A}_i}Q_i(s,a) & \text{with probability} \ 1-\epsilon\\
\mathcal{U}(\mathcal{A}_i)  & \text{with probability} \ \epsilon
\end{cases}
\end{equation}
where $\mathcal{U}(\mathcal{A}_i)$ signifies a sample drawn from the uniform distribution over the action space $\mathcal{A}_i$.

\section{Methodology}
This model employs value-based optimization utilizing the DQN approach within a MARL environment. Throughout the training, agents interact with neighbours, exhibiting cooperative or defective behaviours across various episodes. As illustrated in Fig. 1, each agent $i$ aims to learn a joint policy $\pi_i$ concerning dilemmas and selection actions, informed by their local observations and long-term experiences. A comprehensive explanation of our methodology follows. 
\footnote{Code: \url{https://github.com/itstyren/InteractionMARL-Coop}}

\begin{figure}
\centering
  \includegraphics[width=8.3cm,keepaspectratio]{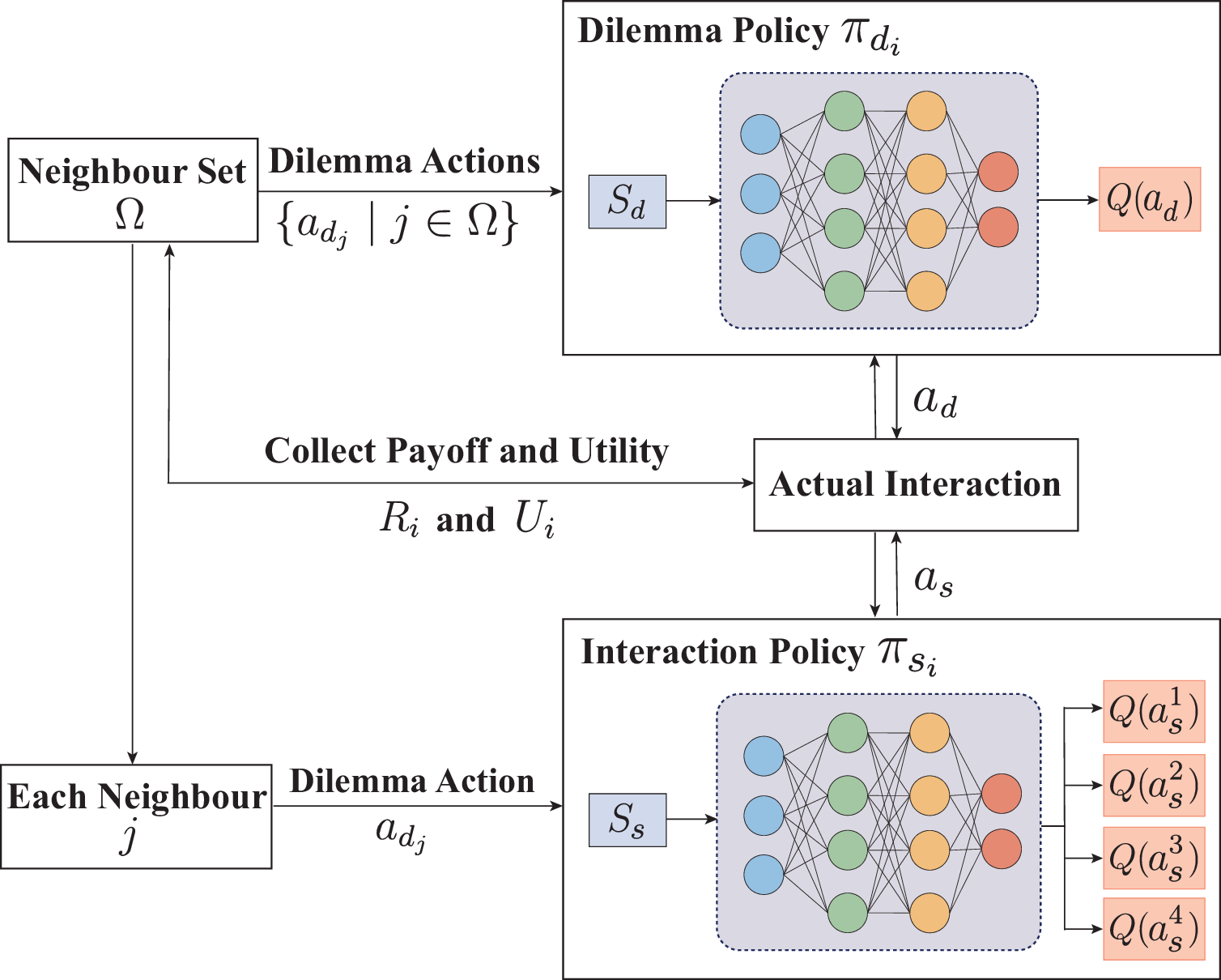}
  \caption{\textbf{Training framework for developing dilemma and interaction strategies.}  Each iteration involves agent $i$ choosing dilemma strategy and selecting neighbouring agents for PDG engagement. Each agent uses two Q-networks: the dilemma policy network, which processes long-term actions in dilemmas by the agent and its neighbours, and the interaction selection network, which assesses neighbours' dilemma actions alongside the agent's previous interactions. The agent calculates the utility of its actions based on the rewards accumulated from past encounters. }
  \label{fig:model_description}
\end{figure}

\subsection{Game Environment}
Agents in our experiment are situated within an $L \times L$ square lattice with periodic boundary conditions. They are placed at specific spatial coordinates and can engage in interactions limited to their von Neumann neighbourhood. The graphical representation employs vertices to denote agents and edges to indicate the relationships between an agent and its four neighbours. The action space for each agent encompasses two strategies: the overall dilemma strategy and the specific interaction selection strategy.  This dual contributes to the formulation of the agent's policy $\pi_i$ expressed as:
\begin{equation}
    \pi_i=(\pi_{s_i},\pi_{d_i})
\end{equation}
where $\pi_{s_i}$ refers to the interaction selection policy dictating whether to interact with its neighbours, while $\pi_{d_i}$ is the dilemma policy that guides the agent in choosing between a cooperative or defective strategy.

\subsubsection{Interaction and Dilemma Actions}
During each timestep of an episode, agents participate in multiple pairwise interactions within the IPD game framework, consisting of two phases: neighbour selection and dilemma strategy determination. In the first phase, agent $i$ chooses game partners through an interaction selection action denoted as $a_{s_i} \in \{0,1\}^4$, informed by previously observed game experiences with its neighbours. This selection action $a_s^i$ is represented by the following bitstring of size 4:
\begin{equation}
    a_{s_i}=(a_{s_i}^1,a_{s_i}^2,a_{s_i}^3,a_{s_i}^4).
\end{equation}

Specifically, a bit value of $1$ for $a_{s_i}^j$ signifies agent $i$ choosing to interact with neighbour $j$, whereas a $0$ implies no interaction. For example, a selection action of $a_s=(1,0,0,1)$ implies interaction only with the first and last neighbours.  Importantly, actual interaction occurs only when both players decide to interact with each other, i.e., when $a_{s_i}^j = a_{s_j}^i = 1$.

In the second phase, each player selects either cooperation or defection as their dilemma strategy. Following the interaction phase, paired co-players engage in one round of PDG pairwise, utilizing the action pair $[a_{d_i}, a_{d_j}]$ where $a_d \in [C, D]$.  Critically, the chosen binary dilemma strategy remains consistent across all interacted neighbours. In other words, a player cannot cooperate with one neighbour while defecting against another neighbour simultaneously. 

\subsubsection{Game Formulation and Reward}
According to their selected dilemma and interaction actions, agent $i$ receives an accumulative payoff at each timestep $t$ by participating in multiple rounds of the PDGs with its currently interacted neighbours, as shown by:
\begin{equation}
r_i(t)=\sum_{j=0}^{n_{i}^{t}}a_{d_{i}}^{T}\mathcal{M}_p a_{d_{j}}
\end{equation}
where $n_{i}^{t}=\sum_{j\in \Omega_i} a^j_{s_{i}}\times a^i_{s_{j}}$ denotes the number of interacted neighbours for agent $i$ at timestep $t$, and $\Omega_i$ is the set of neighbors. In this model, we incorporate learning from previous interactions by employing a weighted moving average of past payoffs~\cite{danku2019knowing}. Thus, the final payoff of agents $i$ at each timestep not just based on the current round's payoff but also includes payoffs from the past $m$ rounds:
\begin{equation}
R_i^{}(t)=\frac{r_{i}^{t}+\sum^M_{m=1}\alpha^mR_{i,m}
}{1+\sum^M_{m=1}\alpha^{m}}
\end{equation}
where $\alpha$ is a parameter that controls the rate of weight decay with increasing $m$, indirectly determining the memory length $M$. To effectively assess the emergent behaviours, $M$ is restricted by the condition $M = \min\{n\mid\alpha^n < 0.01\}$. With $\alpha=0$, agents focus only on the current round, indicating short memory. Conversely, as $\alpha\to1$, agent memory extends to include all previous timesteps, capturing a comprehensive history of experiences.

In EGT research, the Fermi rule is commonly used to model the dynamics of strategy evolution~\cite{szabo1998evolutionary}, reflecting social learning where neighbours tend to imitate the most successful policy observed. A detailed description is elaborated in SI \ref{Supp:EGT}. To adapt these imitation dynamics to the RL context, we have modified the utility function of agent $i$ to align with game payoffs,  which is formulated as follows:
\begin{equation}
U_i(t)=\frac{[\omega_{i}(a_d^{t})+1]R_i(a_{d_i}^{t},a_{s_i}^{t},s_i^{t})-\omega_{i}(\tilde{a}_d^{t})\overline{R}(\tilde{a}_{d}^{t},a_{s}^{t})}{\sum_{a_d\in A_d}\omega_{i}(a_d)+1}
\end{equation}
where $\tilde{a}_{d_i}$ represents a counterfactual dilemma action, condition on the actual action $a_{d}$ taken by agent $i$. The function $\omega_i(\cdot)$ returns the number of neighbours performing a specific action. The term $\overline{R}(\tilde{a}_{d}^{t},a_{s}^{t})$ denotes the average payoff associated with the counterfactual action $\tilde{a}_{d}^{t}$ across the population at timestep $t$. Essentially, the agent raises a retrospective question: ``Would a different past action have led to a more advantageous outcome?" This setting integrates aspects of social learning and RL, allowing agents to compare global information from group about the performance of different actions with their own localized experiences.

\subsection{Training Approach}
In our multi-agent PDG framework, the training methodology aligns with the well-established DQN approach, typically used in single-agent tasks. Our focus is on formulating a joint policy $(\pi_s,\pi_d)$, which utilizes the combined action utilities to calculate the Q-loss for each policy network, guiding both dilemma strategy and interaction selection processes. See SI \ref{Supp:Algorithm} for a detailed elucidation of the training procedure.
\subsubsection{Network Architecture}
Each agent in our independent MARL setup is equipped with a memory buffer, storing experiences from the last $M$ rounds, encompassing a record of both the agent's own actions and those of adjacent agents. The agent updates its memory at each timestep with recent feedback from the local environment, ensuring an accurate representation of its state. The selection and dilemma networks process inputs from long-term neighbour interactions and prior dilemma strategies, respectively. Additionally, both networks consider the dilemma strategies of four neighbouring agents, with actions represented through one-hot encoding and sequenced together.

In the selection phase, agents evaluate the state $s_s\in \mathbb{R}^{2\times 16 \times M}$ , and in the dilemma phase $s_d \in \mathbb{R}^{2\times 5 \times M}$. They operate with two Q-networks configurations, formulating policies independently and without sharing parameters across agents. The architecture of each Q-network includes a dual-layer perception with 32 hidden units and employs the \textit{tanh} activation function for nonlinear transformations.

\subsubsection{Experiment Setup}
During the training stage of our RL experiments, we employ a centralized training with decentralized execution approach~\cite{lowe2017multi}. This method allows agents to access global information regarding the average payoff for potential action in the PDG among the population, thereby facilitating an effective evaluation of their action utilities.

The training involved 900 agents, each equipped with two neural networks, ensuring a broad representation of cooperation dynamics at the population level. To generate experiences for agents, $10$ parallel arenas were established. In each arena during every experimental trial, agents engage in interactions with their neighbours over $6,000$ episodes, each comprising $10$ timesteps, resulting in a total of $60,000$ steps. At the end of each episode, sampled trajectories for agents were aggregated and subsequently forwarded to the respective learner. We compute the gradient by using the Adam optimizer~\cite{kingma2015adam} with a linear annealing schedule of learning rate. To enhance the efficiency of the Q-learner in learning from experience replay, we also implement a common practice known as prioritized experience replay ~\cite{schaul2015prioritized} within the DQN framework. Additional details on hyperparameters, please refer to the SI \ref{Supp:Hyperparameters}.

\section{Results and Discussion}
In this section, we present the outcomes of our experimental investigations, which provide evidence supporting the hypothesis that incorporating the interaction selection action with the dilemma action through RL promotes network reciprocity and cooperation evolution in a spatial PDG setting.

\subsection{Experimental Setup}
Our experiment employs a square lattice setup, randomly assigning agents as cooperators or defectors with equal likelihood. The primary evaluation metric is the fraction of cooperative agents in the population, representing the achieved level of overall cooperation. For robustness, we average the outcomes of the final 10 episodes over the entire training duration. All experiments were replicated five times to ensure replicability. Unless otherwise stated, a memory weight of 0.6 is assigned, incorporating experiences from the previous four rounds as network input.

\subsection{Promotional Effect of RL on Cooperation}

\begin{figure}
\centering
  \includegraphics[width=6.5cm,keepaspectratio]{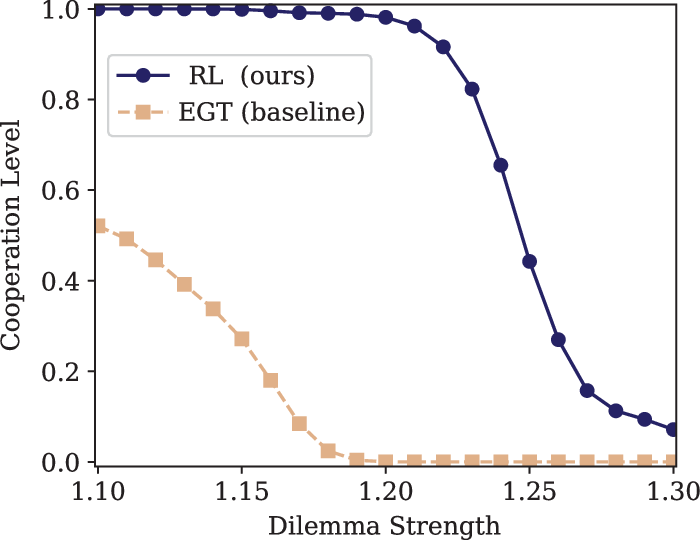}
  \caption{\textbf{RL-based training approach (ours) promotes cooperation more effectively than the EGT (baseline) method.  } The EGT (orange) represents agents solely calculating cumulative payoffs and adjusting dilemma actions through social learning. In contrast, the implementation of effective dilemma and selection policies, guided by RL (blue), has significantly enhanced the level of cooperation within the population.  Our RL-based method maintains full cooperation in the population until dilemma strength exceeds $1.2$. }
  \label{fig: overall_performance}
\end{figure}

To evaluate the effectiveness of our proposed method, we first train a population to learn a combined policy, including both dilemma and interaction strategies under various dilemma strength conditions. Figure \ref{fig: overall_performance} demonstrates that the application of RL in coordinating interaction and dilemma strategies enables the population to sustain a high cooperation level successfully. Notably, this approach proves robust, maintaining its efficacy even in scenarios characterized by increased dilemma intensity. As shown, the MARL system transitions from a complete cooperation phase to a mixed strategy phase exclusively, when the dilemma strength $b>1.2$. 

For comparison, the evolutionary outcomes of conceptually similar models from existing literature are used as a benchmark~\cite{danku2019knowing}. Within EGT framework, agents evaluate equivalent lengths of past payoffs and emulate the most successful dilemma strategy observed in their neighbourhood (a detailed description of EGT methodology, refer to the SI \ref{Supp:EGT}). It is noteworthy that even when the dilemma intensity is reduced to $b=1.1$, only $54\%$ EGT agents opt for cooperative strategies. Moreover, ablation experiments detailed in SI \ref{Supp:ablation} demonstrate that agents utilizing RL to learn dilemma strategies exclusively underperform in comparison to the model proposed in this study. These results suggest that lacking selective interaction in the PDG and the mere imitation of neighbouring successful strategies are insufficient to achieve optimal performance within the population. Additional MARL-based benchmarks are reported in SI \ref{Supp:More_benchmark}.

\subsection{Evolutionary Dynamics of Cooperation}

\begin{figure}
\centering
  \includegraphics[width=8.5cm,height=7.2cm]{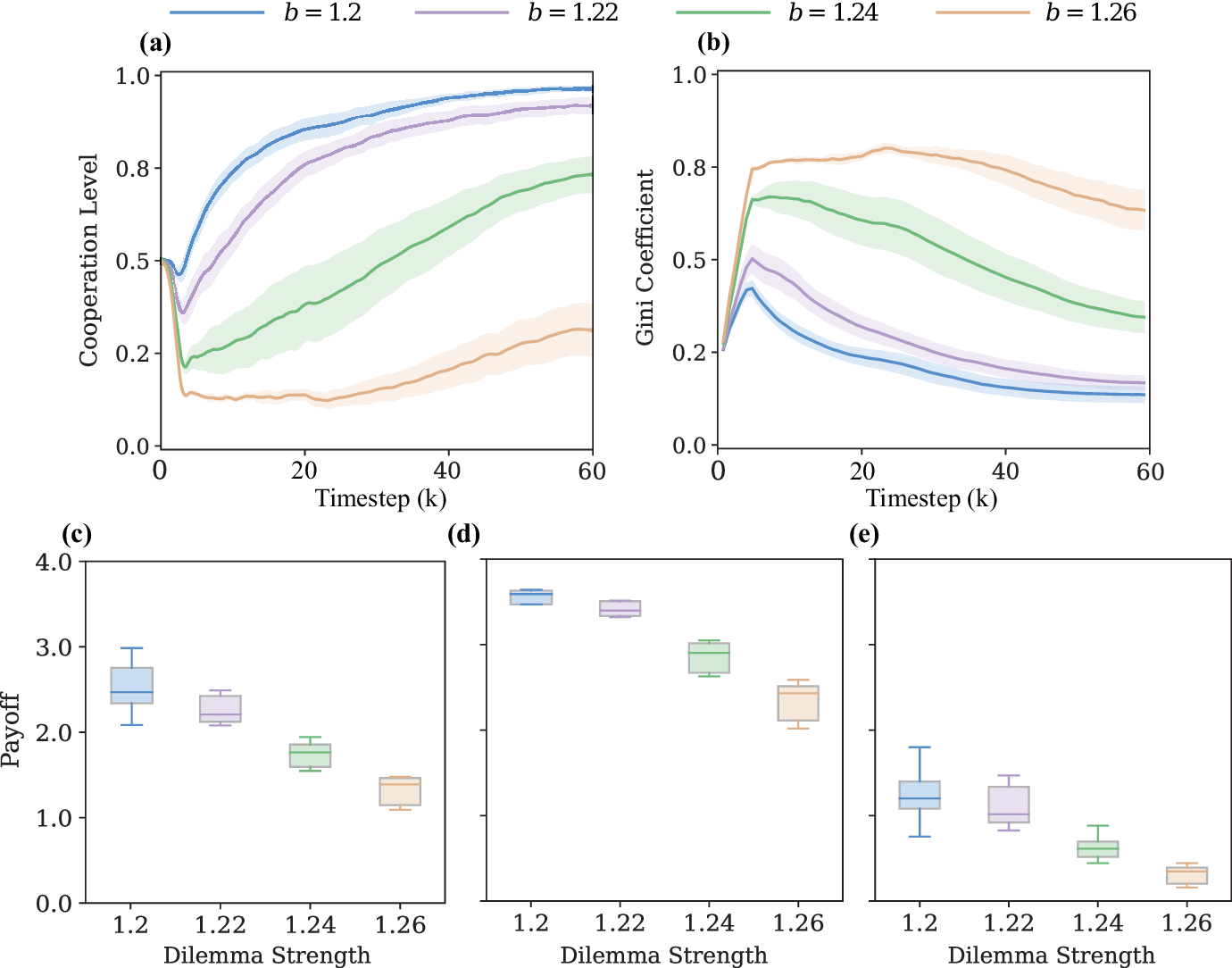}
  \caption{\textbf{The evolution of cooperation and associated payoffs across varying dilemma strengths.} In all scenarios, the fraction of cooperators first decreases and then increases over time, coinciding with reduced average individual payoffs and increased inequality as dilemma strength intensifies. The evaluation encompasses evolutionary trajectories of (a) cooperation level and (b) the Gini Coefficient, alongside metrics including (c) average group payoffs and (d)-(e) payoffs for trained cooperators and defectors, with dilemma strength varying from $b=1.20$ to $1.26$.}
  \label{fig: combine}
\end{figure}

We next investigate the evolutionary dynamics and outcomes of overall dilemma strategies within the population, focusing on the evolution trajectories and average payoffs in four representative dilemma conditions. Figure \ref{fig: combine}(a) reveals a rapid initial decline in the population's cooperation level as the dilemma intensity increases. However, RL agents employing additional interaction strategies exhibit two evident phases in their learning process, aligning with observations from previous EGT experiments~\cite{wang2013insight}: the END period and the EXP period. During the END period, cooperative agents resist the invasion of defectors, and successful cooperators convert those defectors into cooperators in the EXP period. In our experiments, the former period is characterized by a rapid decrease in cooperation levels in the first $3,000$ training step, followed by a phase where these levels rise unless defectors completely dominate in the early stages. Consequently, at the end of the EXP period, the cooperation level in the population significantly decreases, falling from $0.987$ to $0.294$ as $b$ rises from $1.20$ to $1.26$.

The dilemma strength also significantly affects the distribution of individual payoffs, leading to a bifurcated process. As shown in Figure \ref{fig: combine}(b), the group Gini Coefficient exhibits a temporal evolution, initially increasing and subsequently decreasing, hinting at a correlation between payoff equality and the evolution of cooperation.  Notably, a large fraction of cooperators contributes to high levels of group equality. In Figures \ref{fig: combine}(c)-(e), the focus is on evaluating the average payoff for the population, as well as the separate payoffs for cooperative and defective individuals. There is a general decrease in the average payoff as the dilemma increases, which aligns with expectations, given that cooperators are primarily the contributors to the group payoff. Cooperative individuals, however, show greater resilience to tougher dilemma conditions. Specifically, with an increase in dilemma strength from $b=1.20$ to $1.22$, the average payoff per episode decreases from $2.67$ to $2.25$, but this can be lessened by adopting an effective interaction mechanism, resulting in a smaller decrease in cooperative payoff from $3.52$ to $3.41$, compared to a substantial drop for defectors from $1.44$ to $1.10$. This trend indicates the potential of RL agents to develop interaction policies for dilemma scenarios, mitigating adverse effects on their payoff. For detailed information on the average payoff from the trained population, refer to Table \ref{Supp_table:payoff_comparison} in the SI.

\subsection{Efficiency of Learned Interaction Patterns}
\begin{figure}
\centering
  \includegraphics[width=8.3cm,keepaspectratio]{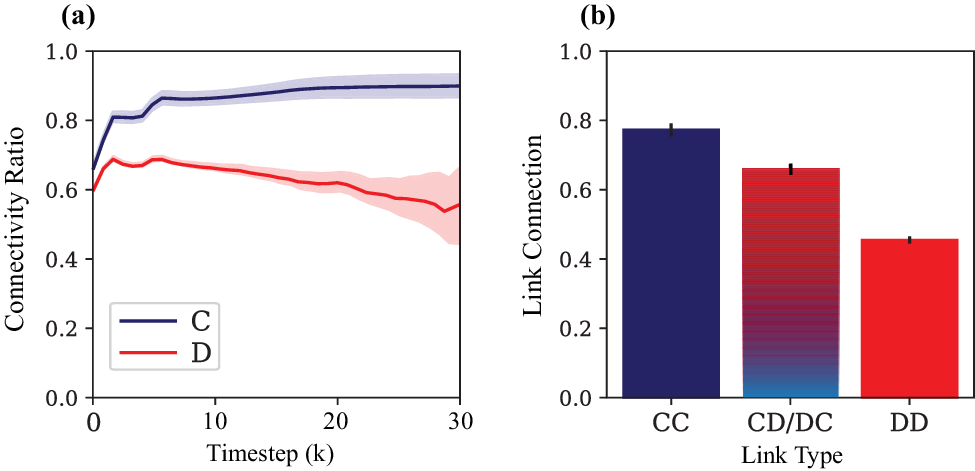}
  \caption{\textbf{Temporal evolution of strategy connectivity and actual link interactions.} RL agents demonstrate enhanced interaction capabilities, increasing connections with cooperative neighbours. (a) The average connectivity ratio for cooperators and defectors in the first half of the total timestep. (b) The frequency of actual link connections between dilemma strategies during the first ten episodes. The dilemma strength is set to $b=1.20$.  }
  \label{fig: interaction}
\end{figure}
\begin{figure}
\centering
  \includegraphics[width=8.4cm,keepaspectratio]{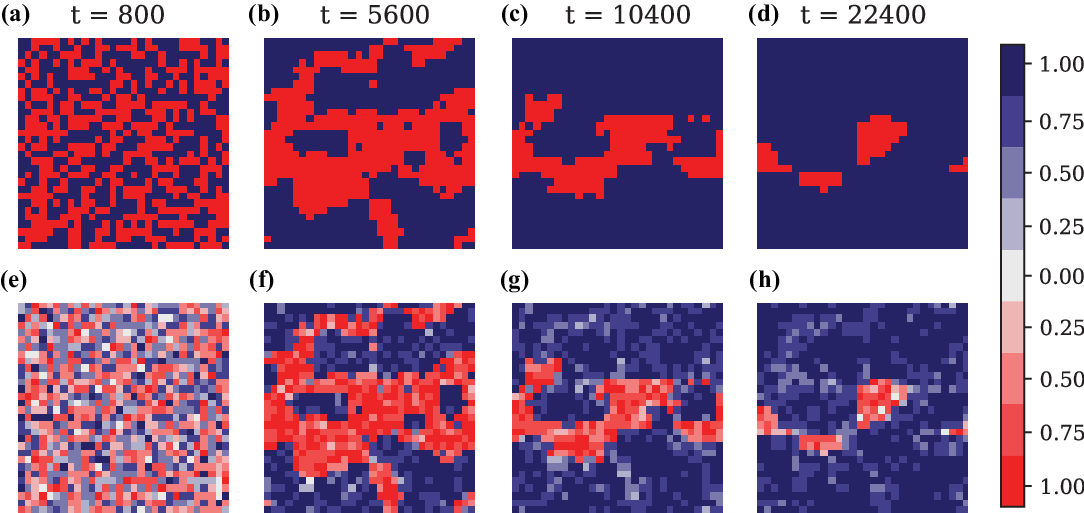}
  \caption{\textbf{Snapshots of the spatial evolution of strategies and their connections.} Cooperative individuals resist defector incursions by forming and expanding clusters. Panels (a)-(d) depict strategy distributions; (e)-(h) illustrate corresponding strategy connections at identical timesteps.  Pixels represent agents as cooperators (blue) and defectors (red), with strategy connectivity ratio varying from 0 (shallow) to 1 (deep). The results are obtained for $b=1.20$. }
  \label{fig: distribution}
\end{figure}

To elucidate the role of interaction selection in network reciprocity and cooperation, we analyze strategy connection and distribution patterns in Figures \ref{fig: interaction} and \ref{fig: distribution}. Through participation in PDG with selective neighbours, RL agents develop policies for distinguishing cooperators from defectors, thereby boosting spatial reciprocity and cooperation. Figure \ref{fig: interaction}(a) depicts how the disparity in average connections between cooperators and defectors increases during the initial half of the total timestep. This leads to a preference for forming connections with cooperators, irrespective of the dilemma strategy chosen by agents.  In Figure \ref{fig: interaction}(b), we evaluate the actual link connections across different link types during the first 10 training episodes, signifying frequency where the chosen co-player reciprocally opts for interaction within the same round. As illustrated, the occurrence of interaction between two neighbouring individuals who both employ the defective strategy (DD link) is merely 45.48\%. In contrast, interactions between two cooperators (CC link) can increase to as high as 77.29\%.  For measurement metrics of agent interactions, see SI \ref{Supp: measurement}.

We next provide intuitive evidence regarding the previously described learned interaction policy and its role in enhancing spatial reciprocity by illustrating the spatial coevolution of the dilemma and interaction strategies within the population over time. Initially, in the END phase depicted in Figures \ref{fig: distribution}(a) and (e), cooperators resist the invasion of defectors by forming small clusters, yet lack an effective neighbour selection strategy. As training progresses, RL agents learn to adapt their interaction strategies in response to neighbouring dilemma strategies, enhancing the influence of spatial reciprocity in promoting the evolution of cooperation. Figures \ref{fig: interaction} and \ref{fig: distribution} (f)-(h) show that, in the EXP phase, individuals within the cooperative cluster show a higher tendency to engage in the PDG with nearby cooperators, while those at the periphery are more likely to avoid interactions with defectors. This selective interaction mechanism favours cooperators and limits the payoff obtained from free-riding behaviours. Consequently, defectors gradually switch their dilemma strategies, leading to the expansion of cooperative clusters.

\subsection{Role of Long-term Experiences}
\begin{figure}
\centering
  \includegraphics[width=8.2cm,keepaspectratio]{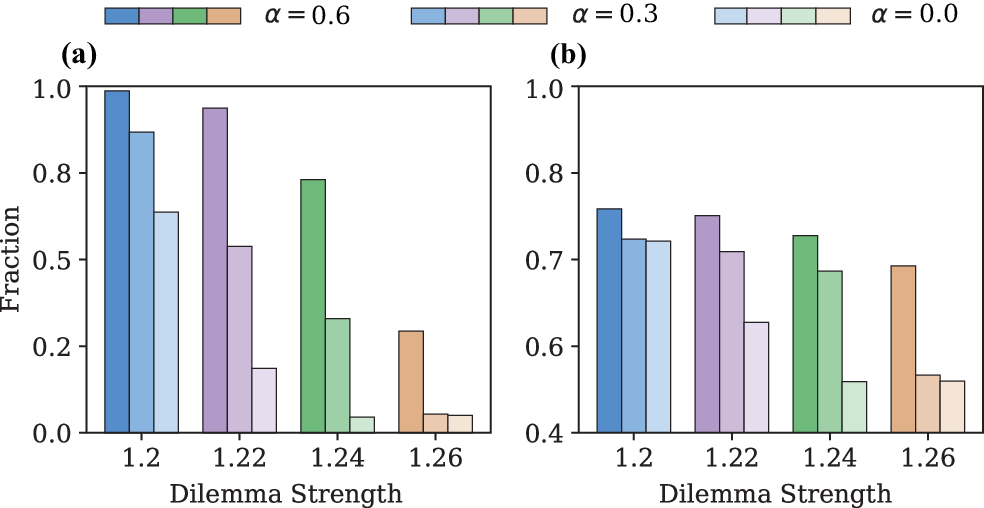}
  \caption{\textbf{Average cooperation level and effective connection for cooperators with varying experience lengths.} Incorporating longer experiences as input enhances group cooperation and cooperative interactions. (a) Post-training cooperation levels in the population. (b) Frequency of cooperators participating in PDG across the initial 20 episodes. These results are obtained for four dilemma strength scenarios, with memory weight $\alpha$ ranging from 0 to 0.6. }
  \label{fig: memory}
\end{figure}
In prior experiments, RL agents use experiences from the last four rounds as network input. Finally, we investigated the influence of varying memory lengths on their learning dynamics, with findings presented in Figure \ref{fig: memory}. Three memory weights were assessed, with $\alpha$ varying from 0.6 to 0, representing input scenarios based on observations from the last, second, and fourth rounds, respectively. Notably, a positive correlation is observed between the group cooperation level and the memory length of trained agents. Figure \ref{fig: memory}(a) illustrates that, for instance, when dilemma strength increased from 1.2 to 1.22,  agents recalling four-time steps maintained group cooperation effectively. Conversely, those relying solely on current information saw a substantial decrease in cooperation levels, dropping from 0.64 to 0.19. 

Furthermore, Figure \ref{fig: memory}(b) echoes the findings of Figure \ref{fig: interaction}, demonstrating that successful cooperation benefits from a preference for efficient communication with neighbouring cooperators. Moreover, the efficiency of interaction selection is also influenced by the length of agent memory. Longer observation inputs are shown to improve the average interaction ratio among cooperators in the END and initial EXP stages, which supports network reciprocity and contributes to the formation and development of cooperative clusters. Interestingly, our findings also reveal that RL individuals consistently and effectively avoid interactions with defectors, irrespective of memory length and dilemma intensity (detailed in SI \ref{Supp: interaction}). Finally, experiments conducted in SI \ref{Supp:Q-Network} demonstrate that the dual Q-network configuration outperforms its single Q-network counterpart, highlighting the advantage of our proposed dual network approach.

\section{Conclusion}
Our computational model enables agents to interact selectively with their neighbours and protect cooperative behaviours from antisocial influences by MARL and iterative trial-and-error in simulations. Unlike existing RL cooperation studies, our approach does not rely on predefined social norms or external incentives~\cite{ueshima2023deconstructing,tennant2023modeling}. By integrating a spatial PDG into the training environment, we extend focus from paired interactions to those within a spatial structure, enabling a deeper analysis of how long-term learning impacts the coevolution dynamics of cooperation and selective interaction.

Our findings reveal that trained agents are capable of differentiating between neighbouring cooperators and defectors by observing information from their immediate surroundings, which enhances the network reciprocity and the associated group cooperation. Consistent with theoretical models~\cite{szolnoki2020blocking}, our findings suggest that the performance of the learned interaction mechanism in promoting cooperation is attributed to its ability to help populations form homogeneous strategic clusters. These clusters are crucial for resisting invasions by defectors, especially in the early stages of development. Additionally, we confirm a positive correlation between agent memory length and the effectiveness of interaction selection, which in turn, aids the evolution of cooperative behaviors~\cite{park2022cooperation}.

In conclusion, we emphasize the significance of understanding the learning dynamics in interaction selection and their contribution to fostering cooperation. These insights offer new insights into the emergence of cooperation within social dilemmas. Moreover, the computational framework developed here has broader implications, providing a versatile tool for investigating and examining mechanisms behind the evolution of cooperation in spatially structured environments. Integrating psychological complexity in our training framework emerges as a promising direction for future research. Understanding these mechanisms may provide solutions to social dilemmas and strengthen cooperation within both human societies and artificial intelligence systems.

\newpage
\appendix 



\section*{Acknowledgments}
The authors would like to thank anonymous reviews for their constructive comments. We also appreciate the insightful discussions with Xiaomei Mi and Yang Li related to this paper, and are grateful Haotong Zhang and Runqi Zhao for their support with computational resources. Finally, we acknowledge the assistance given by Research IT and the use of the Computational Shared Facility at The University of Manchester.

\bibliographystyle{named}
\bibliography{reference}


\end{document}


\maketitle

\appendix 
\section{Implementation Details}  \label{Supp:implementation}
\subsection{Training Procedure} 
 \label{Supp:Algorithm}
For a comprehensive understanding, Algorithm ~\ref{alg: algorithm} details the training procedure for multi-agent environments in the spatial Prisoner's Dilemma Game (PDG).

\begin{algorithm}[H]
    \caption{Multi-Agent Training in the Spatial PDG}\label{alg: algorithm}
    \begin{algorithmic}
    \FOR{each episode $e = 1$ to $M$} 
        \STATE Observe initial states $s_{s}$ and $s_{d}$ for each agent
        \FOR{timestep $t=1$ to max-episode-length}
            \FOR{each agent $i$ to $N$}
            \STATE Select interaction action $a_{s_i} \sim \pi_{s_i}(s_{s_i})$
            \STATE Select dilemma action $a_{d_i} \sim \pi_{d_i}(s_{d_i})$
            \ENDFOR
            \STATE Execute joint action $A=(a_1,\dots,a_n)$, observe reward $r$ and new state $s'$ for all agents
            \FOR{each agent $i$ to $N$}
            \STATE Calculate the final payoff $R_i$ as: 
            \begin{equation*}
            R_i^{}=\frac{r_{i}+\sum^M_{m=1}\alpha^mR_{i,m}
            }{1+\sum^M_{m=1}\alpha^{m}}
            \end{equation*}
            \STATE Calculate the utility of joint action $U_i$ as:
            \begin{equation*}
            U_i=\frac{[\omega_{i}(a_d)+1]R_i(a_{d_i},a_{s_i})-\omega_{i}(\tilde{a}_d)\overline{R}(\tilde{a}_d)}{\sum_{a_d\in A_d}\omega_{i}(a_d)+1}
            \end{equation*}
            \FOR{interaction and dilemma action $a$}
                \STATE Store transition  $(s_a,a,U,s'_a)$ in replay buffer $\mathcal{D}_a$ 
                \STATE Randomly sample minibatch of transitions
                \STATE Perform gradient descent on the loss function: 
                \begin{equation*}
                L(\theta_{a_i})=(u+\gamma\max_{a'}\overline{Q}(s',a')-Q(s,a))^2
                \end{equation*}
            \ENDFOR
            \STATE Periodically update target network parameters:
            \begin{equation*}
                \theta_{a_i}'\leftarrow \tau\theta_{a_i}+(1-\tau)\theta_{a_i}'
            \end{equation*}
            \ENDFOR
        \ENDFOR
    \ENDFOR
    \end{algorithmic}
\end{algorithm}

\subsection{Hyperparameters}  \label{Supp:Hyperparameters}
In all experiments conducted, the reward discount factor, $\gamma$, was consistently set at $0.99$. The learning rate for the Adam optimizer followed a linear decay pattern, commencing at an initial rate of $1.0$ and gradually decreasing to a final rate of $0.05$. For each experimental iteration, training was carried out using five unique random seeds. This training was executed in parallel across ten distinct arena environments.  Concerning policy exploration, the rates, symbolized as $\epsilon$, for the dilemma and interaction scenarios exhibited a linear reduction from $1$ to $0.05$ and from $1$ to $0.1$, respectively, over the initial $2,000$ timesteps.

Throughout the training phase, a replay buffer with a maximum storage capacity of $10,000$ samples was employed. Updates to the network parameters were executed following every addition of a batch comprising five samples into the replay buffer, using mini-batches of size $32$. Additionally, the target network experienced soft updates at each timestep, implementing a Polyak averaging approach with a coefficient $\tau$ set at $ 0.01$. The prioritization exponent, denoted as $\alpha$, was consistently maintained at a value of $0.6$. Concurrently, the importance-sampling correction factor, $\beta$, underwent a linear progression from $0.4$ to $1$.

\subsection{Evaluation of Agent Interactions} \label{Supp: measurement}
In our experiments, we evaluate the effect of the emergent interaction selection strategy on several outcome metrics. Here, a comprehensive explanation is provided for each outcome metric utilized in the main manuscript. In accordance with the main text, we denote the total number of agents in the population as $N$ and use $\Omega$ to represent the set of neighbouring agents.
\begin{itemize}
\item  \textbf{Connectivity Ratio ($CR$)}: This metric reflects the propensity of agents to establish connections with either cooperators or defectors within the network. For a given agent $i$, the Connectivity Ratio is formally defined as the ratio of the number of neighbouring agents that are connected to it relative to the total number of potential connections within agent $i$'s neighbourhood.
\begin{equation}
CR_i=\frac{\sum_{j\in\Omega_i}a_{s_j}^i}{|\Omega_i|}.
\end{equation}
\item  \textbf{Effective Connection ($EC$)}: This metric measures the average effective interaction strength of different strategies. For a given agent $i$, the Effective Connection calculates the proportion of successful PDG with neighbours when central agent $i$ adopts a specific strategy.
\begin{equation}
EC_i=\frac{\sum_{j\in\Omega_i}a_{s_i}^j \times a_{s_j}^i}{|\Omega_i|}.
\end{equation}
\item  \textbf{Link Connection ($LC$)}: This metric quantifies the frequency of actual interactions occurring between different types of linked strategies within a given population.  It serves as a measure of the effectiveness of interconnections between various dilemma strategies. We categorize these interconnections based on the nature of the linkages, resulting in a division into three types: cooperator-cooperator (CC), cooperator-defector (CD/DC) and defector-defector (DD). 
\item \textbf{Link Proportion ($LP$)}: This metric is designed to analyze the proportion of different link types within a network. It aims to provide a quantitative measure of the relative frequency of each link type in the context of the group's dynamics. It offers a percentage representation of each edge category within the overall network structure, classified based on the interaction types (CC, CD/DC and DD). 
\item \textbf{Gini Coefficient ($Gini$)}: This metric quantifies the degree of inequality in a distribution, specifically the payoff distribution in our study.  It is quantified as a value ranging from 0 to 1, where 0 denotes perfect equality (every individual possesses identical income) and 1 denotes perfect inequality (a single individual holds all the income). When the payoffs for all agents are arranged in ascending order, with each payoff $r$ assigned a rank $i$, it is calculated using the following formula:
\begin{equation}
Gini=\frac{\sum_{i=1}^n(2i-n-1)r_i}{n\sum_{i=0}^n r_i}.
\end{equation}
\end{itemize}

\section{Description of EGT model}  \label{Supp:EGT}
In the context of Evolutionary Game Theory (EGT) models within structured populations,  the application of the Monte Carlo simulation procedure is commonly executed using a random sequential strategy for updating processes. Initially, all competing strategies are distributed uniformly at random on a square lattice, which is characterized by periodic boundary conditions. The implementation of the PDG on a square lattice can be described as follows. Firstly, one selected player obtains its payoff $R_i$ by summing up all the payoffs accrued from interactions in the PDG with all its neighbouring players. Subsequently, one nearest neighbour is chosen at random, and this player similarly calculates its overall payoff $R_j$ from its adjacent interactions. Finally, player $i$ attempts to impose its strategy $s_i$ onto player $j$, with the probability given by the Fermi function:
\begin{equation}
W(s_i\to s_j)=\frac{1}{1+\exp{[(R_{s_i}-R_{s_j})/K]}}
\end{equation}
where $K=0.1$ represents the uncertainty factor in the strategy selection process. In the $K\to0$ limit, player $i$ will imitate the strategy of player $j$ of and only if $R_i>R_j$, indicating a deterministic imitation based on superior payoffs. In contrast, as $K\to+\infty$, the imitation process becomes entirely random. In our experiments, setting $K=0.1$ implies that strategies yielding higher payoffs are predominantly imitated, albeit with a few exceptions allowing for some variability in strategy adoption. By repeating the aforementioned basic steps $N$ times in a single Monte Carlo step, each player gets the opportunity to update their strategy once on average.

\section{Differentiating Our RL from EGT Settings}  \label{Supp:distinctions_EGT_RL}
The aim of this study is to promote cooperative behaviours among RL agents through network reciprocity and partner selection. Prior research in EGT highlights the coevolution of strategies and interaction relationships in enhancing group cooperation. This section aims to outline key distinctions between our RL framework and traditional EGT approaches.

First, our RL model enables agents to directly observe the long-term actions of both themselves and their neighbours, thus enhancing adaptability and fostering robust cooperation and zero-shot social coordination within structured networks. Second, our agents operate without initial knowledge, developing high-order norms~\cite{santos2018social} through experiences, which enables them to spontaneously establish interaction behaviours that stabilize network reciprocity, echoing findings from EGT literature~\cite{van2010learning}. Additionally, our approach combines trial-and-error learning with counterfactual reasoning to refine reward functions, improving policy development and partner selection efficiency beyond EGT. Finally, the epsilon parameter in the deep Q-network (DQN) encourages exploration and prevents premature convergence, facilitating a balanced learning process.

\section{Additional Result}  \label{Supp:additional_result}
\subsection{Comparison between Individual Payoff} \label{Supp:payoff_comparison}

\begin{table*}[t]
  \centering
    \begin{tabular}{lrrrrrrrrrrrrrrr}
        \toprule
        \multicolumn{1}{c}{} & \multicolumn{3}{c}{Mean} & \multicolumn{3}{c}{Median}  & \multicolumn{3}{c}{Std} & \multicolumn{3}{c}{Min} & \multicolumn{3}{c}{Max} \\
        \cmidrule(rl){2-4} \cmidrule(rl){5-7} \cmidrule(rl){8-10} \cmidrule(rl){11-13} \cmidrule(rl){14-16}
          Dilemma & {Ep.} & {Coop.} & {Def.} & {Ep.} & {Coop.} & {Def.} & {Ep.} & {Coop.} & {Def.} & {Ep.} & {Coop.} & {Def.} & {Ep.} & {Coop.} & {Def.}\\
        \cmidrule(r){1-1} \cmidrule(rl){2-2} \cmidrule(rl){3-4}  \cmidrule(rl){5-5} \cmidrule(rl){6-7} \cmidrule(rl){8-8} \cmidrule(rl){9-10} \cmidrule(rl){11-11}  \cmidrule(rl){12-13} \cmidrule(rl){14-14}  \cmidrule(rl){15-16} 
          $b=1.20$ & 2.67 & 3.52 & 1.41 & 2.52 & 3.59 & 1.33 & 0.41 & 0.16 & 0.35 & 2.15 & 3.18 & 0.96 & 3.48 & 3.64 & 2.40 \\
          $b=1.22$ & 2.25 & 3.41 & 1.10 & 2.21 & 3.40 & 1.01 & 0.15 & 0.08 & 0.22 & 2.08 & 3.32 & 0.83 & 2.49 & 3.52 & 1.47 \\
          $b=1.24$ & 1.74 & 2.86 & 0.62 & 1.76 & 2.90 & 0.61 & 0.13 & 0.16 & 0.12 & 1.55 & 2.63 & 0.44 & 1.94 & 3.05 & 0.88 \\
          $b=1.26$ & 1.33 & 2.34 & 0.31 & 1.39 & 2.43 & 0.35 & 0.15 & 0.21 & 0.10 & 1.09 & 2.02 & 0.16 & 1.48 & 2.59 & 0.44 \\
        \bottomrule
    \end{tabular}
  \caption{\textbf{Average payoffs for trained population and distinct cooperative versus defective strategies.}The presence of an effective interaction mechanism reduces the effect of higher dilemma intensity on the payoffs of cooperative individuals compared to those who defect.}
  \label{Supp_table:payoff_comparison}
\end{table*}

The data presented in Table \ref{Supp_table:payoff_comparison} illustrates a comparative analysis of the payoffs accrued by a trained population utilizing various cooperative and defective strategies across four distinct levels of dilemma strength. As indicated by the data, increased dilemma strength diminishes the average payoff of the whole population as well as cooperators. However, cooperators consistently gain greater benefits than defectors. This indicates that cooperative agents are capable of learning an efficient interaction mechanism to mitigate the negative impact of heightened dilemma strength on their payoffs.

\subsection{Emergent Strategies of Interaction Selection} \label{Supp: interaction}

\begin{figure}[t]
\centering
  \includegraphics[width=6.5cm,keepaspectratio]{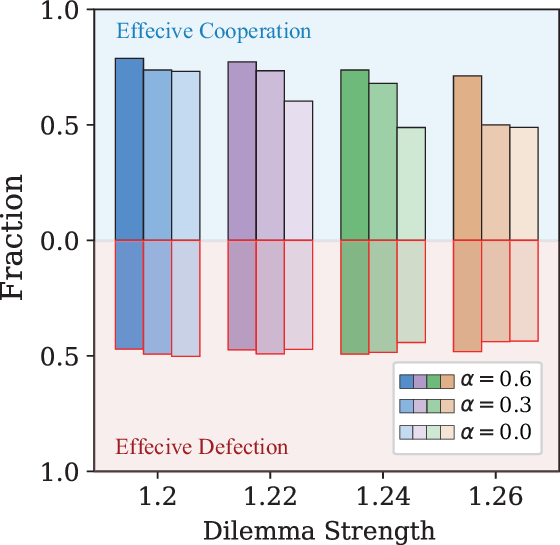}
  \caption{\textbf{Frequency of effective neighbour interaction for cooperators and defectors across the initial 25 episodes.} Cooperative agents consistently exhibit greater effectiveness in engaging PDG compared to defective agents. Notably, effective defection maintains at a low level, regardless of memory length and dilemma intensity. }
  \label{sup_fig: effective_interaction}
\end{figure}

\begin{table*}
  \centering
    \begin{tabular}{lrrrrrrrrrrrr}
        \toprule
        \multicolumn{1}{c}{} & \multicolumn{3}{c}{CR.C} & \multicolumn{3}{c}{CR.D}  & \multicolumn{3}{c}{EC.C} & \multicolumn{3}{c}{EC.D}\\
        \cmidrule(rl){2-4} \cmidrule(rl){5-7} \cmidrule(rl){8-10} \cmidrule(rl){11-13} 
          Dilemma & {M1} & {M2} & {M4} &  {M1} & {M2} & {M4} & {M1} & {M2} & {M4}&  {M1} & {M2} & {M4} \\
        \cmidrule(r){1-1} \cmidrule(rl){2-2} \cmidrule(rl){3-4}  \cmidrule(rl){5-5} \cmidrule(rl){6-7} \cmidrule(rl){8-8} \cmidrule(rl){9-10} \cmidrule(rl){11-11} \cmidrule(rl){12-13}
          $b=1.20$ & 0.86 & 0.88 & 0.90 & 0.77 & 0.75 & 0.55 & 0.74 & 0.75 & 0.80 & 0.51 & 0.50 & 0.46 \\
          $b=1.22$ & 0.78 & 0.87 & 0.91 & 0.70 & 0.77 & 0.64 & 0.60 & 0.74 & 0.79 & 0.46 & 0.50 & 0.46\\
          $b=1.24$ & 0.74 & 0.83 & 0.86 & 0.61 & 0.79 & 0.77 & 0.50 & 0.68 & 0.74 & 0.42 & 0.49 & 0.50 \\
          $b=1.26$ & 0.61 & 0.75 & 0.85 & 0.53 & 0.61 & 0.80 & 0.49 & 0.50 & 0.72 & 0.41 & 0.42 & 0.49 \\
        \bottomrule
    \end{tabular}
  \caption{\textbf{The average connection ratio and  the efficacy of connections within populations employing cooperative and defective strategies.} Enhancing the length of experiences within the input network can augment the connectivity associated with cooperative strategies and elevate the interaction efficiency for cooperators. The results evaluated three scenarios, characterized by memory lengths of 1, 2 and 4, across various dilemmas strength.}
  \label{tab:strategy_interaction}
\end{table*}

\begin{table*}[t]
  \centering
   \resizebox{\textwidth}{!}{
    \begin{tabular}{lrrrrrrrrrrrrrrrrrr}
        \toprule
        \multicolumn{1}{c}{} & \multicolumn{3}{c}{LC.CC} & \multicolumn{3}{c}{LC.CD/DC}  & \multicolumn{3}{c}{LC.DD} & \multicolumn{3}{c}{LP.CC} & \multicolumn{3}{c}{LP.CD/DC} & \multicolumn{3}{c}{LP.DD}\\
        
        \cmidrule(rl){2-4} \cmidrule(rl){5-7} \cmidrule(rl){8-10} \cmidrule(rl){11-13} \cmidrule(rl){14-16} \cmidrule(rl){17-19} 
          Dilemma & {M1} & {M2} & {M4} &  {M1} & {M2} & {M4} & {M1} & {M2} & {M4}&  {M1} & {M2} & {M4} &  {M1} & {M2} & {M4}  &  {M1} & {M2} & {M4}\\
        \cmidrule(r){1-1} \cmidrule(rl){2-2} \cmidrule(rl){3-4}  \cmidrule(rl){5-5} \cmidrule(rl){6-7} \cmidrule(rl){8-8} \cmidrule(rl){9-10} \cmidrule(rl){11-11} \cmidrule(rl){12-13}  \cmidrule(rl){14-14} \cmidrule(rl){15-16} \cmidrule(rl){17-17} \cmidrule(rl){18-19}
          $b=1.20$ & 0.83 & 0.76 & 0.79 & 0.61 & 0.61 & 0.63 & 0.49 & 0.47 & 0.43 & 0.52 & 0.87 & 0.98  & 0.08 & 0.05 & 0.01 & 0.40 & 0.08 & 0.01\\
          $b=1.22$ & 0.76 & 0.81 & 0.78 & 0.55 & 0.61 & 0.61 & 0.48 & 0.47 & 0.45 & 0.13 & 0.61 & 0.97 & 0.07 & 0.07 & 0.02 & 0.80 & 0.32 & 0.02\\
          $b=1.24$ & 0.51 & 0.79 & 0.81 & 0.49 & 0.57 & 0.62 & 0.47 & 0.48 & 0.47 & 0.01 & 0.27 & 0.63 & 0.06 & 0.11 & 0.10 & 0.92 & 0.61 & 0.27 \\
          $b=1.26$ & 0.59 & 0.56 & 0.78 & 0.53 & 0.51 & 0.49 & 0.60 & 0.47 & 0.46 & 0.02 & 0.01 & 0.57 & 0.08 & 0.06 & 0.09 & 0.90 & 0.93 & 0.34 \\
        \bottomrule
    \end{tabular}}
  \caption{\textbf{The link connection and the proportion of different link types within a network.} The adaptive selection of interacted neighbour within the population contributes to an increase in the effectiveness of interconnections among cooperators. Simultaneously, it leads to a reduction in the intensity of interactions between defectors. }
    
  \label{tab:link_interaction}
\end{table*}
Building upon the discussion of the emergent interaction selection strategy presented in the main text, we provide a more detailed analysis here. As shown in Figure \ref{sup_fig: effective_interaction}, it is evident that the fraction of effective interactions between cooperative individuals within the group and their neighbours consistently exceeds that of defective individuals. This disparity in interaction dynamics serves as a protective mechanism, shielding cooperative members from potential exploitation by free riders. Furthermore, there is a notable correlation between the effectiveness of cooperative interactions and the length of experience utilized as input in the Q-network by agents. It is crucial to highlight, however, that the average effectiveness of interactions associated with defection strategies remains below 50\%, irrespective of the variations in individual memory length.  This implies that during the early stages of evolution, agents inherently prioritize the identification and subsequent disconnection from defectors. Such a strategy facilitates the formation of cooperative clusters, which are instrumental in resisting the infiltration of defectors in the END stages of the evolutionary process.

In table \ref{tab:strategy_interaction} and \ref{tab:link_interaction}, we present the statistical features derived from our experiments. The results indicate that the effectiveness of emergent interaction strategies for RL agents is positively correlated with the length of experience that can be incorporated into the network. For instance, as the dilemma strength intensifies from $b=1.20$ to $b=1.26$, the proportion of effective cooperative connections decreases from $0.86$ to $0.61$ when information from a single round is considered. This reduction is confined to a mere $0.05$ when the network input includes data from four rounds of historical experiences. However, at high dilemma intensities, an excessively long memory length also enhances the interaction effectiveness among individuals employing defective strategies. This indicates that RL agents are capable of learning effective interaction with their neighbours only within certain bounds of dilemma strength. Consequently, the introduction of additional mechanisms becomes necessary when the dilemma strength is increased.

\subsection{Dilemma Behavior Ablation Study} \label{Supp:ablation}
\begin{figure}
\centering
  \includegraphics[width=6.2cm,keepaspectratio]{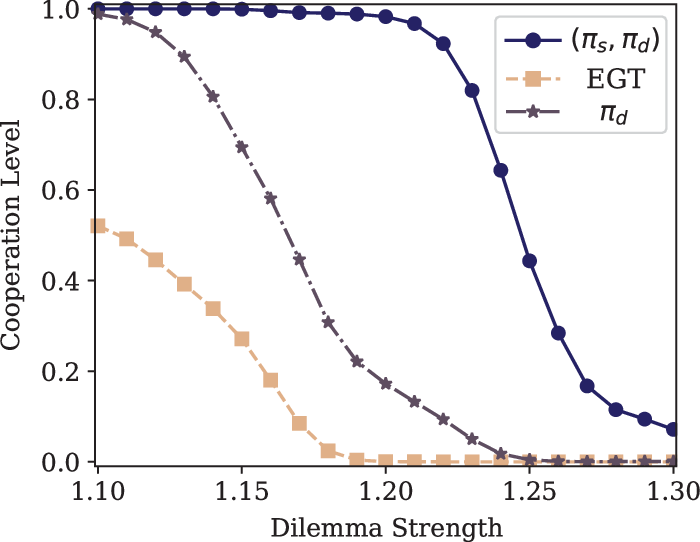}
  \caption{\textbf{Comparative Performance of the RL and EGT Model.} RL can promote cooperation more effectively than traditional EGT methods, and this effect is further enhanced by incorporating learned neighbour selection.} 
  \label{sup_fig: different_models}
\end{figure}
We conduct a series of ablation experiments to investigate the necessity of simultaneously learning dilemma strategies in conjunction with neighbour selection. Figure \ref{sup_fig: different_models} provides the performance of the RL model when employing simultaneous learning of dilemma strategies with interactions, in contrast to using learning dilemma strategies in isolation, and includes a comparative analysis with the EGT baseline. The results demonstrate that an RL agent, even when only learning dilemma strategies from scratch, can achieve a higher level of cooperation than agents engaging in social learning. For instance, agents that learn dilemma strategies through RL transition to a phase of full defection only when the dilemma strength reaches $b=1.25$, thus underscoring the importance of self-learning.  

Additionally, the effectiveness of RL is further augmented when neighbour selection dynamics are introduced. As shown, RL agents engaged in simultaneous learning of both dilemma and interaction strategies demonstrate the ability to maintain a full cooperation state under increased variations in dilemma strength. 

\subsection{Training Using Single or Dual Q-Networks} \label{Supp:Q-Network}
\begin{table}
  \centering
    \begin{tabular}{lrrrrrrr}
        \toprule
        \multicolumn{1}{c}{}  & \multicolumn{3}{c}{Interaction} & \multicolumn{3}{c}{Payoff}     \\
        \cmidrule(rl){2-4} \cmidrule(rl){5-7} 
          Model & {CR.C} & {CR.D} &  {EC.C} & {Ep.} & {Coop.} & {Gini.}  \\
        \cmidrule(r){1-1} \cmidrule(rl){2-3}  \cmidrule(rl){4-4} \cmidrule(rl){5-6} \cmidrule(rl){7-7}   
          Dual & 0.90 & 0.55 & 0.80 & 2.67 & 3.52 & 0.11  \\
          Single& 0.81 & 0.67 & 0.67 & 2.85 & 2.85 & 0.17 \\
        \bottomrule
    \end{tabular}
  \caption{\textbf{Performance Differences Between a Single DQN and a Dual DQN Configuration.}  A separate configuration enhances the efficiency of group interactions and increases the average payoff for cooperative agents. The dilemma strength is set to $b=1.20$. }
  \label{tab:dual_single}
\end{table}
In the main manuscript, each agent is allocated two Q-networks that represent its decision-making process concerning actions in dilemmas and interactions. Here, we compare the performance of the adopted dual Q-network approach with a methodology employing a singular Q-network to represent both types of actions simultaneously. For a single network outputting actions, the output action for agent $i$  is denoted by a five-bit string:
\begin{equation}
a_i=(a_{d_i},a_{s_i}^1,a_{s_i}^2,a_{s_i}^3,a_{s_i}^4)
\end{equation}

The configuration of the Q-network remains consistent with the main text, including the dual-player perception with 32 hidden units and the use of the $tanh$ activation function for nonlinear transformations. 

We observe that with each group member using a single Q-network, full cooperation is achieved at a dilemma intensity of 1.2, surpassing the $0.98$ cooperation level of the dual network setup. However, the dual network configuration outperforms the single network setup in encouraging cooperation. Table \ref{tab:dual_single} provides a comprehensive analysis of the dual versus single Q-network frameworks. As shown, agents employing dual networks for dilemma strategies and neighbour interactions learn more effective neighbour selection mechanisms, leading to cooperators with higher average connectedness ($0.90$ vs. $0.81$) and more effective interaction ($0.80$ vs. $0.67$) within the population. Additionally, while the average group payoff is slightly higher with a single network, employing dual networks notably enhances the level of cooperation and reduces inequality within the group, as evidenced by a decrease in the Gini coefficient from 0.17 to 0.11. This indicates that segregating dilemma behaviour and interactive behaviour into distinct networks allows RL agents to avert suboptimal strategies and better protect cooperator interests.

\begin{table}[h]
  \centering
    \begin{tabular}{lrrrrrrr}
        \toprule
        \multicolumn{1}{c}{}  & \multicolumn{3}{c}{Cooperation Level} & \multicolumn{3}{c}{Episode Payoff}     \\
        \cmidrule(rl){2-4} \cmidrule(rl){5-7} 
          Method  & {1.0} & {1.1} &  {1.2} & {1.0} & {1.1} & {1.2}  \\
        \cmidrule(r){1-1} \cmidrule(rl){2-4} \cmidrule(rl){5-7}    
          Ours $(\pi_s,\pi_d)$ & \textbf{1} & \textbf{1} & \textbf{0.98} & \textbf{4} & \textbf{3.96} & \textbf{2.67}  \\
          Ours $(\pi_d)$ & \textbf{1} & 0.99 & 0.18 & \textbf{4} & 3.33 & 1.59  \\
          SVO & 0.41 & 0.22 & 0.16 & 1.68 & 0.97 & 0.69 \\
          Selfishness & 0.45 & 0.24 & 0.16 & 1.83 & 1.03 & 0.63 \\
        \bottomrule
    \end{tabular}
  \caption{\textbf{Comparative performance of ours and two Recent MARL-based approaches.} Agents trained using our proposed MARL framework demonstrate superior performance in terms of overall cooperation levels and average episode payoff across various strengths of dilemmas.}
  \label{tab:more_benchmark}
\end{table}

\subsection{Comparisons with More Benchmark} \label{Supp:More_benchmark}
Finally, we evaluate our proposed training framework by comparing it with two contemporary RL methods within the same spatial social dilemma environment. Specifically, we chose SVO~\cite{mckee2020social} and Selfishness~\cite{roesch2024selfishness} approaches as MARL-based benchmarks for this comparative analysis. Both our proposed method and the selected benchmarks employ payoff modifications based on the comparison of individual versus group performance to foster prosocial preferences among players. The key distinction is our approach lies in guiding policy updates by adopting a form analogous to the Fermi function while incorporating both long-term returns and interactions. This approach allows for a more nuanced adaptation of strategies, emphasizing the balance between individual strategy and dynamic interactions among agents. Results from Table \ref{tab:more_benchmark} suggest that learning effective interaction mechanisms aligned with dilemma strategies can induce better cooperative outcomes.

\bibliographystyle{named}
\bibliography{SI_reference}